\documentstyle[11pt,nature]{article}

\lefthead{Lee et al.}
\righthead{}

\begin{document}
\author{\raggedleft To appear in November 4th issue of Nature}
\title{Multiple stellar populations in the globular cluster $\omega$ Centauri 
            as tracers of a merger event\\}
\author{\raggedright Y.-W. Lee$^{\star}$, J.-M. Joo$^{\star}$, Y.-J. Sohn$^{\star}$, S.-C. Rey$^{\star}$, 
H.-c. Lee$^{\star}$ \& A. R. Walker$^{\dagger}$}
\affil{\raggedright$^{\star}$Center for Space Astrophysics, Yonsei University, Seoul 120-749, Korea\\
       \raggedright$^{\dagger}$National Optical Astronomy Observatories/Cerro
Tololo Interamerican Observatory (NOAO/CTIO), Casilla 603, La Serena, Chile}
%\affil{}

{\bf  The discovery of the Sagittarius dwarf galaxy$^{1}$, which is being tidally
disrupted by and merging with the Milky Way, supports the view that the halo of the
Galaxy has been built up at least partially by the accretion of similar dwarf systems.
The Sagittarius dwarf contains several distinct populations of stars$^{2,3}$, 
and includes M54 as its nucleus, which is the second most massive globular cluster
associated with the Milky Way. The most massive globular cluster is $\omega$ Centauri,
and here we report that  $\omega$ Centauri also has several distinct stellar populations,
as traced by red-giant-branch stars. The most metal-rich red-giant-branch stars are
about 2 Gyr younger than the dominant metal-poor component, indicating that $\omega$
Centauri was enriched over this timescale. The presence of more than one epoch of 
star formation in a globular cluster is quite surprising, and suggests that
$\omega$ Centauri was once part of a more massive system that merged with the Milky Way,
as the Sagittarius dwarf galaxy is in the process of doing now. Mergers probably
were much more frequent in the early history of the Galaxy and $\omega$ Centauri
appears to be a relict of this era.}
\\
\\

        As part of our investigation of the luminosity-metallicity relation of the
RR Lyrae stars in the globular cluster $\omega$ Cen, we have obtained 2K $BV$
CCD frames with the CTIO 0.9-m telescope that cover 40$\times$40 arcmin$^{2}$ in a
3$\times$3 grid centred on the cluster, and covering out to approximately half the
tidal radius. In total, 40 - 42 frames were taken in each filter and each field.
The seeing was between 1.0 and 1.7 arcsec, and all of the observing nights
(April 5 - 10 1996) were fully photometric. The $B$ and $V$ magnitudes of individual 
stars were measured with the point-spread-function (PSF)-fitting programs
DAOPHOT II and ALLSTAR in the standard manner$^{4}$. As a by-product of this
investigation, we obtained high-quality homogeneous $BV$ colour-magnitude data
for more  than 130,000 stars in the field toward $\omega$ Cen, which represents the
most extensive photometric survey to date for this cluster.

        Figure 1  shows a $V$ versus $B-V$ colour-magnitude  diagram (CMD)  for 
stars in our  programme field. We note the presence of  several distinct 
red-giant-branches (RGB)s with a red, presumably metal-rich, sequence well separated 
from other bluer metal-poor ones. This
feature was not evident in previous photometry$^{5}$ with smaller sample sizes and
larger photometric uncertainties. The radial distribution of the most metal-rich RGB stars
is not significantly different from those of the metal-poor ones, which confirms that
they belong to $\omega$ Cen. Note also that
the red-clump that must be  associated with the most metal-rich RGB  is clearly
apparent, partially overlapping the metal-poor RGBs.  The presence of other interesting
features on the CMD, such as the blue-tail phenomenon of  the horizontal-branch
(HB) and the blue straggler stars, illustrates the diversity of stellar populations in
this cluster. The signature  of field-star  contamination is also evident,  primarily 
as a swathe of stars with 0.3 $\lesssim$ $(B-V)_{o}$ $\lesssim$ 1.2.  These stars will belong
to the foreground galactic disk population.

        In order to further investigate  the discrete nature of the  RGB, we have
plotted in Figure 2 a histogram of the distribution  of colour difference
between  each  RGB  star and the RGB fiducial of the  most  metal-poor
component. The RGB stars are  selected in a relatively  narrow magnitude range
12.4 $<$ $V$ $<$ 12.9, so that  the field-star  and red-clump  contamination is
minimized, and which also avoids artificial mixing in the histogram stemming from
metallicity dependence  of the RGB  slope. The presence  of several  distinct RGBs is
confirmed, although the most metal-rich  component is not as well distinguished as
in Figure 1 because of the small sample size of such stars in this magnitude range.

        Figure 3  shows our  population models. These illustrate the relative age
estimation from the location of the red clump associated with the most
metal-rich RGB with respect to blue HB stars associated with the most
metal-poor component. Panel (a) shows the case where all stars have the same age
despite their different metallicities, which produces a red HB that is clearly  much
bluer than the observed red-clump confined within the  two most extreme
RGBs. Panel (b) shows our best match with  the observed CMD in Figure 1, which
suggests that the most metal-rich population  in $\omega$ Cen is some  2 Gyr younger
than the most metal-poor population in this system. This internal  age-metallicity
relation is clear evidence that $\omega$ Cen has enriched itself over this timescale.

        The multimodal  metallicity distribution  function and  the age-metallicity
relation found in this study suggest that the protocluster  of $\omega$ Cen was
massive enough to undergo some self-enrichment and several early bursts of star
formation. The relatively extended enrichment period of about 2 Gyr then indicates
that the  initial  evolution of $\omega$ Cen occurred  away  from the  dense  central
regions of the young  Galaxy, for if this  were not the case,  one would expect
the gaseous material to have  been stripped from the  cluster on a much shorter
timescale$^{8}$. The  plausible resolution   of this
problem is that $\omega$ Cen has evolved within a dwarf-galaxy-sized gas-rich subsystem
(the Searle and Zinn$^{9}$ fragment) until  it merged with and disrupted by our  Galaxy
some 2 Gyr after the  formation of its first-generation metal-poor stars, leaving
its core as today's globular cluster $\omega$ Cen. This is consistent with the fact  that
the multiple populations  observed in $\omega$ Cen bear a strong resemblance  to the
Sagittarius dwarf system, where  three distinct RGBs are identified with an internal
age-metallicity relation that spans more  than 3 Gyr (refs 2,3). Alternatively, it  might be
argued that a  merger of  several clusters could  also create  a composite object
with multiple populations. The well  defined peaks in the  metallicity distribution
function in Figure 2, however, require a merger of  at least four clusters, and the
possibility of this occurring may be  extremely low.  Furthermore, cluster  mergers could  only
have occurred in dwarf galaxies with  velocity dispersions much smaller than  the
Milky Way's$^{10}$. Thus,  even if $\omega$ Cen was formed by cluster
mergers, it is probably  still a relic of  a dwarf galaxy  subsequently accreted by
our Galaxy. If our interpretation is  correct, the case of $\omega$ Cen  and that of the
Sagittarius dwarf  system would  provide direct  evidence for  past and continuing 
accretion of protogalactic fragments,  which suggest that similar  accretion events
may have  continued  throughout the   history of  Galactic formation.  Extensive
photometric surveys  for other   massive globular clusters,   especially those with
peculiar CMD  morphology, such  as M22  with broad  RGB like  $\omega$ Cen,  and
NGC 2808   with bimodal  HB  distribution,  will undoubtedly   help to  reveal
whether they also represent the relics of the Galaxy building blocks.

\clearpage
\newpage

\acknowledgments
{\raggedright {\bf Acknowledgements.} Support for this work was provided by the Creative Research
Initiatives Program of the  Korean Ministry of Science \& Technology, and  in part,
by the Korea Science \& Engineering Foundation. S.-C.R. was a visiting astronomer at
CTIO/NOAO, which is operated by the Association of Universities for Research in Astronomy, Inc., 
under cooperative agreement with the National Science Foundation.}

{\raggedright Correspondence should be addressed to Y.-W. L. (e-mail: ywlee@csa.yonsei.ac.kr)}.

\clearpage
\newpage
\begin{figure}
 \caption{Colour-magnitude  diagram of  50,129 stars  in the  direction of $\omega$ Cen. 
These diagrams were obtained from a mosaic of nine 2K CCD fields. Only stars with at least 20
detections and small photometric errors ($\sigma_{V}$ $<$ 0.05 mag and  
$\sigma_{B-V}$ $<$ 0.071 mag) have been  plotted. 
Panel (a) is for all  stars in  our programme  field, while  panel (b)  is only  for stars  
located between 2.58$^{\prime}$ and 15.48$^{\prime}$ from the cluster centre. There are 
several distinct RGBs and a  red-clump associated with  the most metal-rich  component.
Two RGB loci from the new Yale isochrones$^{6}$ (Z = 0.0004, 0.005) are also
compared in panel (b) that bracket the metallicity range of $\omega$ Cen.}
\end{figure}

\begin{figure}
 \caption{Histogram of the distribution  of colour difference.
Colour difference between each RGB star and the RGB fiducial of the most metal-poor
component, $\Delta$$(B-V)$, is plotted in the range 12.4 $<$ $V$ $<$ 12.9.
The metallicities (Z values) of four distinct RGBs are also marked.}
\end{figure}

\begin{figure}
 \caption{Stellar population models. This models illustrate the estimation of age 
difference between
the red-clump associated with the most metal-rich  (Z = 0.005)  RGB and  the blue HB
associated with  the most metal-poor (Z = 0.0004) component.
Dots and crosses are individual HB stars from synthetic HB models$^{7}$ and the solid lines are
from new Yale isochrones$^{6}$. Panel (a) is for the case that all stars have the 
same age despite their different metallicities, while panel (b) is for the case that the
most metal-rich population is 2 Gyr younger than the most metal-poor population.
Only  $\Delta$$t$ of about  2 Gyr reproduces the features on the observed CMD in Fig. 1.}
\end{figure}

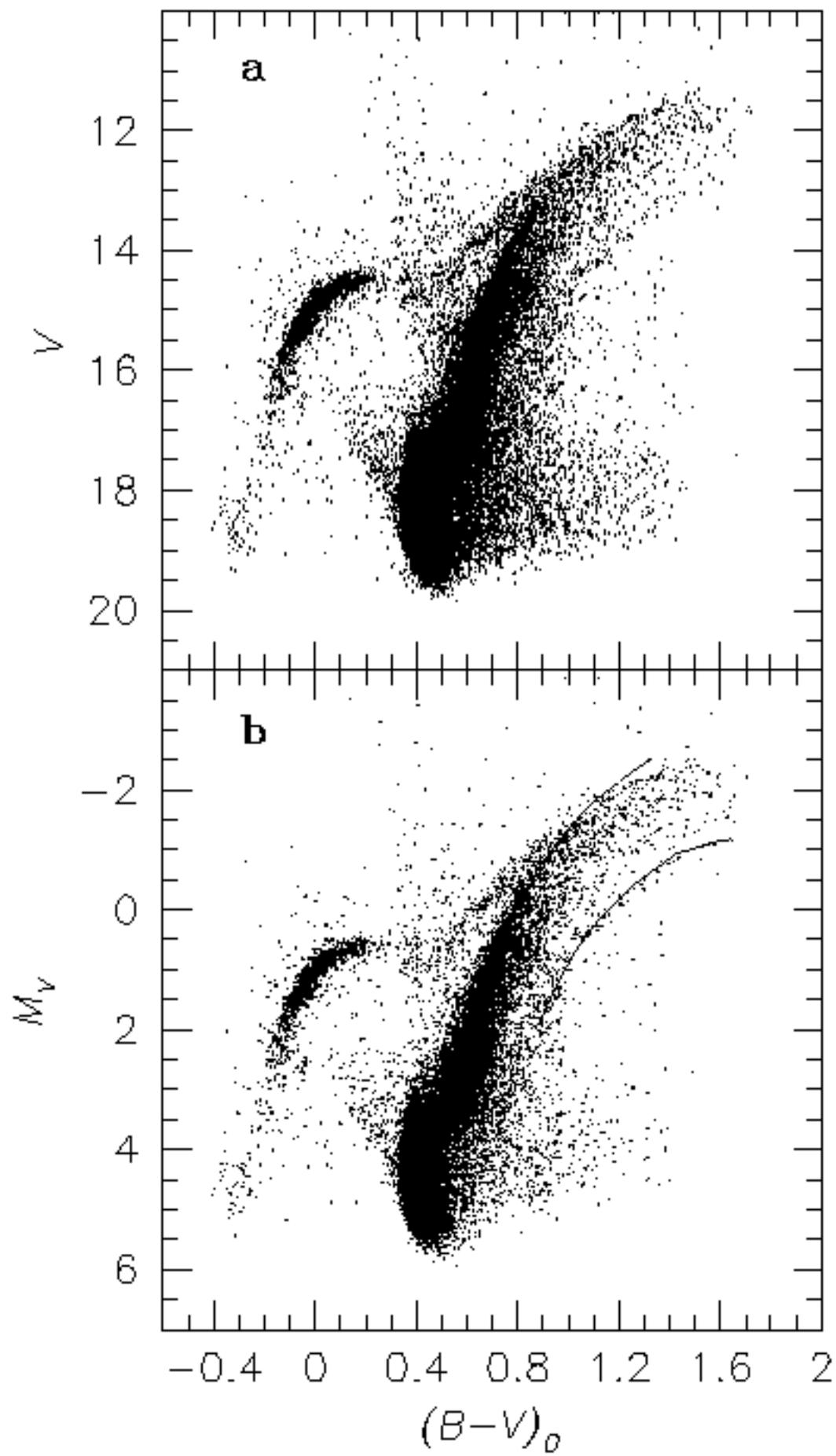
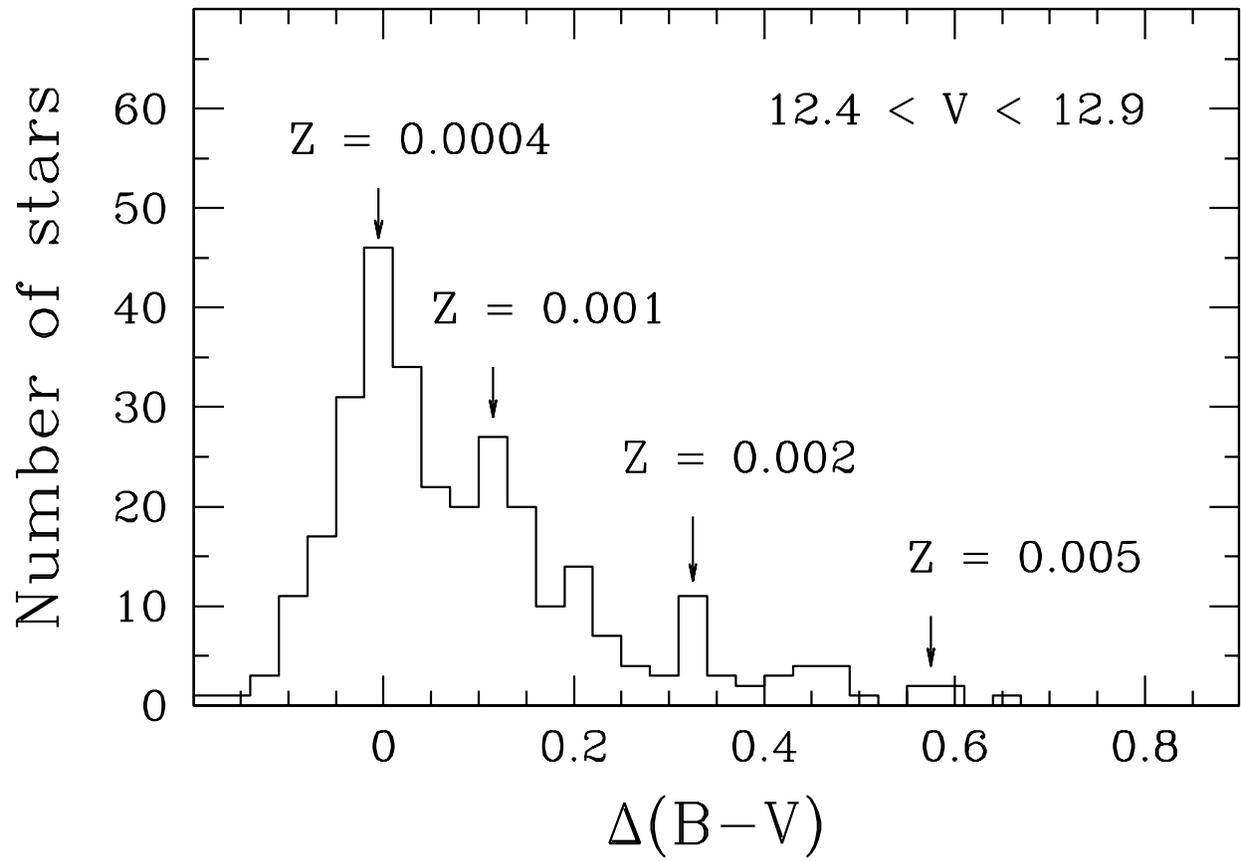
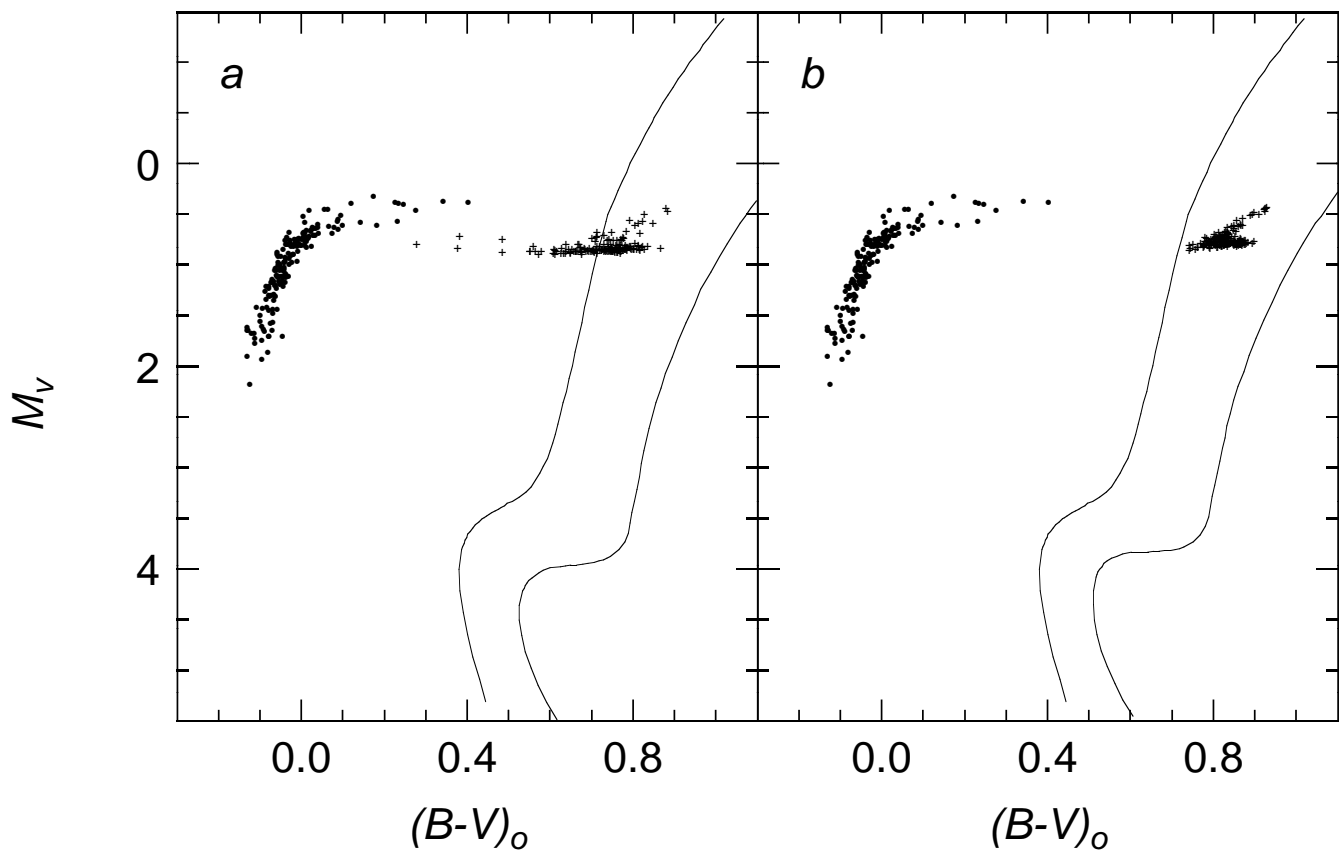
\end{document}